\documentstyle[prd,aps,epsf]{revtex}
\begin{document}

\input epsf \renewcommand{\topfraction}{0.8}
\twocolumn[\hsize\textwidth\columnwidth\hsize\csname
@twocolumnfalse\endcsname

\title{Tiling with almost-BPS junctions.}
\author{P. M. Saffin}
\address{DAMTP,
            Silver Street,
            Cambridge.\\ 
            E-mail: p.m.saffin@damtp.cam.ac.uk
}
\date{\today}
\maketitle
\begin{abstract}
In the light of recent studies of BPS triple junctions in the
Wess-Zumino model we describe techniques to construct infinite lattices
using similar junctions. It is shown that whilst these states
are only approximately locally BPS they are nevertheless stable
to small perturbations, giving a stationary tiling of the plane.
\end{abstract}
\vskip2pc]



Domain walls have found their way into many areas of physics,
ranging from the small scale in solid state laboratories, 
where they can appear as crystal dislocations,
to the large scales of cosmology. Such objects can
come about when there are vacuum states which are disconnected 
yet degenerate in energy, although one can envision walls appearing
when the distinct vacua have different energy, but then the walls
are not static. 

The case of interest here is supersymmetric field theory, where
the distinct vacua come about from a
polynomial superpotential. As we
shall see one is able to construct states where two or more walls
meet at a junction to create a configuration which saturates a Bogomol'nyi
bound. These junctions have been investigated recently in the context
of the Wess-Zumino model by two groups \cite{gibbons99}
\cite{carroll99}, pointing out that these junctions preserve $\frac{1}{4}$
of the \mbox{$N=1$} supercharges.
One may also motivate this study from the viewpoint of supersymmetric
QCD. There it is found that gluino condensates can form, leading to
the effective degrees of freedom satisfying a Wess-Zumino model
\cite{taylor83}. In this case distinct vacua are present, 
giving domain walls which are BPS states \cite{decarlos99}.

An intriguing possibility then arises due to the existence of
junctions - one may make a network of 
domain walls in a similar manner to string networks 
\cite{sen98}. Here however we shall see that these networks of
domain walls are only locally approximately BPS states rather than
full BPS states. In doing this we shall come across a wonderful
variety of patterns, including some of the Euclidean tilings
\cite{grunbaum}.

Although a connection is not clear, such arrangements are familiar
in fluid mechanics \cite{edwards94} and the physics of granular layers
\cite{melo95}. In these Faraday experiments the fluids are driven by
an external force which can generate instabilities, such as convection
instabilities. A consequence of this is the generation of
diverse cell patterns, from regular to quasi-patterns.


We shall follow the approach of Gibbons and Townsend 
in our choice of model \cite{gibbons99},
specifically the  model under investigation is the bosonic sector of the
Wess-Zumino model, reduced to 2+1 dimensions. 
Much of this section can be found in their work but is
reproduced here for completeness.
The Lagrangian is defined by,
\begin{eqnarray}
{\cal L}&=&\frac{1}{4}\partial_\mu\bar{\phi}\partial^\mu\phi-|W'(\phi)|^2.
\end{eqnarray}
For static configurations we may use the energy density to derive
a Bogomol'nyi equation. This is facilitated by the introduction of
the complex coordinate \mbox{$z=x+iy$}, whereupon the energy density
becomes,
\begin{eqnarray}
{\cal H} = \left|{\partial\phi\over\partial z} - 
e^{i\alpha}{\overline {W'}}\right|^2
+ 2{\rm Re}\left( e^{-i\alpha}{\partial W\over \partial z}\right)
+ {1\over2} J(z,\bar z).
\end{eqnarray}
Here we have introduced an arbitrary phase $\alpha$, and $J(z,\bar z)$
is defined by,
\begin{eqnarray}
J(z,\bar z) =
\left({\partial\phi\over\partial\bar z}\,
{\partial\bar\phi\over\partial z} - {\partial\phi\over\partial z}\,
{\partial\bar \phi\over\partial\bar z}\right)\, .
\end{eqnarray}
We may derive a Bogomol'nyi bound by defining the quantities,
\begin{eqnarray}
Q&=&\frac{1}{2}\int {\rm d}x\rm{d}y\; {\it J}(z,\bar z),\\
T&=&2\int {\rm d}x\rm{d}y\; \frac{\partial {\it W}}{\partial z},
\end{eqnarray}
leading to
\begin{eqnarray}
E =\int {\rm d}x\rm{d}y\; {\cal H}\ge {\it Q} +  |{\it T}|\, , 
\end{eqnarray}
which is saturated by solutions of the first order equation,
\begin{eqnarray}
\label{bogeqn}
{\partial\phi\over\partial z} = e^{i\alpha}{\overline {W'}}\, .
\end{eqnarray}
We now focus on models where the scalar field potential energy
density, $|W'(\phi)|^2$, contains isolated, degenerate minima.
This allows for the presence of domain walls, in particular if there
are more than two minima there will be more than one type of domain
wall, giving the possibility of a wall junction. 
The question of what type of walls exist was investigated in
\cite{abraham91}, showing that not all 4D field theories with more 
than two disconnected vacua admit junction solutions, in contradiction
to a statement made in \cite{carroll99}.
We note here that an existence proof for a class of junction solutions
of the second order equations has been provided in \cite{bronsard96}.

Junctions may be thought of as the meeting point of a number of domain
walls. Each domain wall interpolates between two vacua and one may
associate a complex topological charge to them \cite{abraham91},
\begin{eqnarray}
\label{tension}
T_{ab}&=&2e^{iarg(W(\phi_b)-W(\phi_a))}|W(\phi_b)-W(\phi_a)|,
\end{eqnarray}
with BPS walls having a tension, $\mu_{ab}=|T_{ab}|$. This formula
for the tensions will be of great use later, as the wall tensions
are needed to describe the form of  the junctions.

\begin{figure}[!htb]
\centerline{\setlength\epsfxsize{80mm}\epsfbox{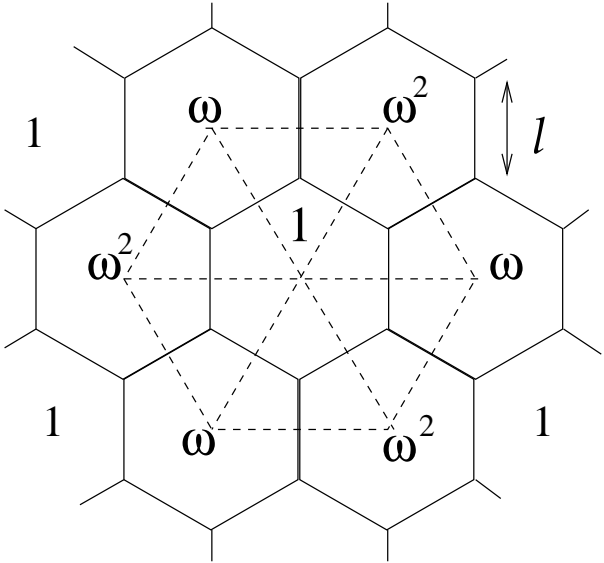}}
\caption{Schematic plot of a hexagonal array.}
\label{hex_pic}
\end{figure}

To be specific we now choose a superpotential which leads
to three distinct vacua, placed where $\phi$ is a cube root of unity,
(1, $\omega$, $\omega^2$),
\begin{eqnarray}
W(\phi)&=&\phi-\frac{1}{4} \phi^4.
\end{eqnarray}
In this case the tensions are all equal, leading to a triple junction
with 120$^\circ$ separating the sectors. One possible lattice 
using these junctions consists
of hexagonal domains, pictured in Fig. \ref{hex_pic}.
One may ask the question whether such a network could be a BPS state 
\cite{gibbons99}\cite{carroll99},
that this is not the case may be seen from two perspectives. 
Firstly we note
that the lattice can at best be perturbatively stable, meaning it
is not the lowest energy state for the given boundary data,
and so is not BPS. If one of the
domains were to tunnel to a different vacuum,
liberating an energy of approximately $3l\mu$, the network would not
recover its original form; the newly tunnelled phase would propagate through 
the entire lattice.
Secondly, one may try to create such 
a network by solving the BPS equations in
the plaquettes of the dual lattice
(the dotted triangles in Fig. \ref{hex_pic}), 
then gluing the plaquettes together.
This would clearly
lead to a network of BPS junctions. However we note that each junction
has a winding associated to it and that the three junctions it
connects to have the opposite winding, with $\phi$ being proportional
to $z$ or $\bar{z}$ ($z$ here is measured from the centre of the junction.)
We see then that while one junction may satisfy (\ref{bogeqn}) its
three neighbours satisfy an anti-BPS relation,
\begin{eqnarray}
\label{antibogeqn}
{\partial\phi\over\partial \bar{z}} = e^{i\beta}{\overline {W'}}\, .
\end{eqnarray}
A BPS junction would therefore
be connected to three anti-BPS junctions, so no global coordinate
system exists that could make the whole lattice BPS. We also note that
such a construction would lead to discontinuities in the derivatives
of $\phi$ where the 
dual plaquettes  meet because one is solving different equations in each of
them.

To establish whether a network could exist we performed a numerical
simulation of the second order Lagrange equations and searched for a hexagonal
structure, using a lattice with 
periodic boundary conditions. Once a tiling had been found we tested
its stability against local fluctuations 
by randomly perturbing the field, with the expected
result; static lattices exist so long as the domain sizes are greater
than the width of the walls. An example is given below in
Fig. \ref{hex}.
There remains, however, the possibility of non local instabilities
other than tunnelling. In the thin wall limit the walls approach
the BPS limit, and then one may expand or shrink a plaquette
without changing the angles of the junctions, so keeping the energy
the same. These non local zero modes do not survive outside the
thin wall limit. As a plaquette is shrunk the walls which make up its
boundary deviate more from the BPS limit, so increasing there tension,
causing the plaquette to further collapse. This has been tested
numerically, finding that such modes were not excited by the local
random fluctuations but did occur when the initial data contained
one plaquette smaller than the rest.
\begin{figure}[!htb]
\centerline{\setlength\epsfxsize{80mm}\epsfbox{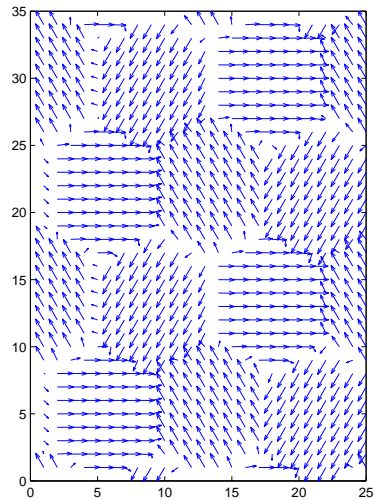}}
\caption{Plot of a hexagonal array using the full field equations.
The arrows represent the complex field $\phi$ withe their size and
direction corresponding to the magnitude and phase respectively.}
\label{hex}
\end{figure}
An interesting property of the domain walls in this model is that
the wall interpolating between two vacua is effected by the other
vacua, making the the field trace out a curve in $\phi$ space which
is not straight. An even more remarkable property of the BPS domain
walls is that these curves are straight when plotted in the
superpotential, W, plane \cite{fendley90}\cite{carroll99}. 
To see this consider a domain wall which
is independent of $y$, so that \mbox{$z\rightarrow x$} in
(\ref{bogeqn}). Then multiplying both sides of (\ref{bogeqn}) by
$\frac{\partial \bar{\phi}}{\partial x}$ yields
\begin{eqnarray}
|\frac{\partial \phi}{\partial x}|^2&=&e^{i\alpha}\frac{\partial
W}{\partial x},
\end{eqnarray}
where $\alpha$ is now found to be the argument of the topological
charge on the wall \cite{abraham91}. This shows that the imaginary
component of $e^{i\alpha}W(\phi)$ is a constant, leading to BPS domain
walls tracing out straight lines in the $W$ plane. This is illustrated
below in Fig. \ref{config}, where the hexagonal array of
Fig. \ref{hex} has been mapped to the $\phi$ and $W$ planes.
The density of dots in this figure represents the volume of physical
space occupying that region of field space, the cusps (vacua)
are the most dense, as we would expect for most of the field 
being in a vacuum. The straight lines joining the vacua in
Fig. \ref{config} (b) confirm the almost-BPS nature of the domain walls.
\begin{figure}[!htb]
\centerline{\setlength\epsfxsize{80mm}\epsfbox{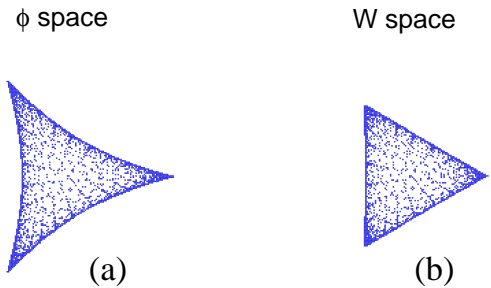}}
\caption{Plot of the configuration space covered by the hexagonal
network. (a) is $\phi$ space and (b) $W$ space.}
\label{config}
\end{figure}

\begin{figure}[!htb]
\centerline{\setlength\epsfxsize{80mm}\epsfbox{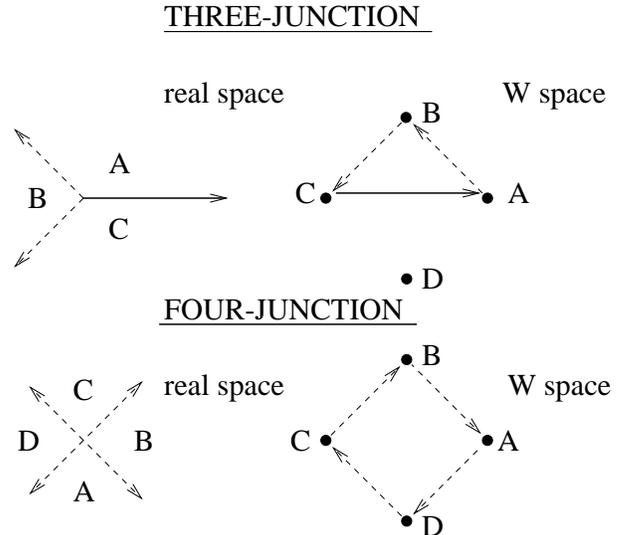}}
\caption{The different vertices allowed for the quintic
superpotential,
A, B, C and D are the locations of the vacua.}
\label{quad_vert}
\end{figure}

\begin{figure}[!htb]
\centerline{\setlength\epsfxsize{80mm}\epsfbox{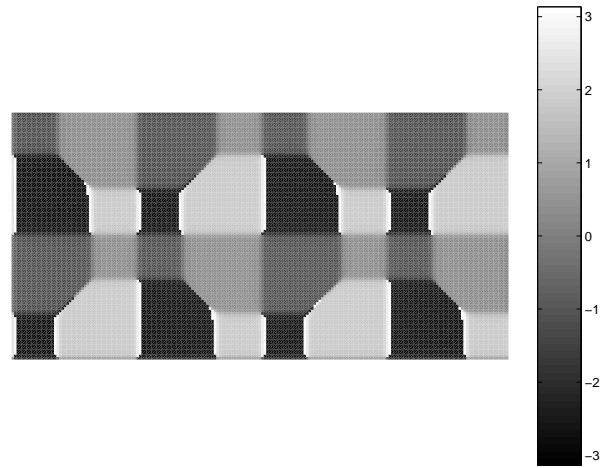}}
\caption{Plot of a octagonal array, indicating the phase of the
different domains.}
\label{oct2_phase}
\end{figure}

\begin{figure}[!htb]
\centerline{\setlength\epsfxsize{80mm}\epsfbox{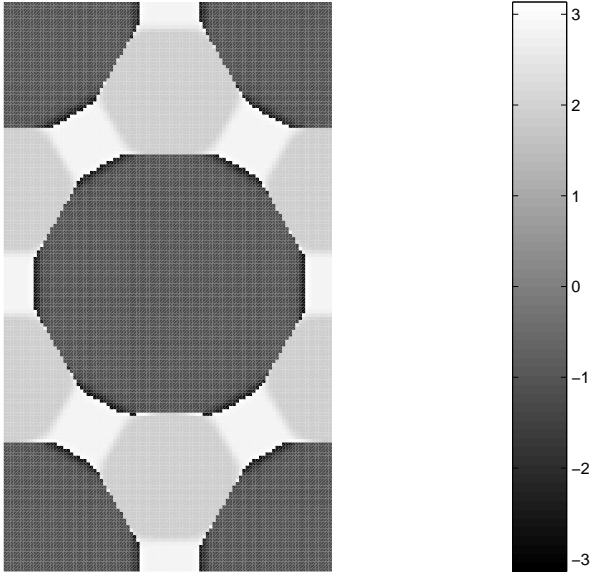}}
\caption{Plot of a dodecagon array, indicating the phase of the
different domains.}
\label{dod_phase}
\end{figure}

\begin{figure}[!htb]
\centerline{\setlength\epsfxsize{80mm}\epsfbox{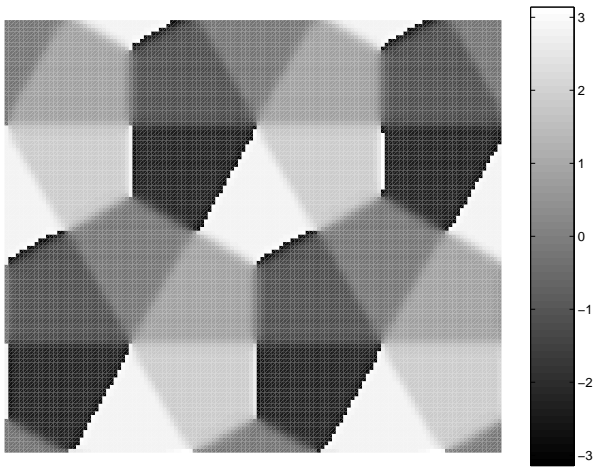}}
\caption{Plot of a hexagonal array, indicating the phase of the
different domains.}
\label{hex_sex_phase}
\end{figure}

Using what we have learned above we may now be more adventurous in the
choice of superpotential. Using the quintic superpotential,
\mbox{$W(\phi)=\phi-\frac{1}{5} \phi^5$}, we find that the vacua for $\phi$
are the fourth roots of unity. In this case there are six domain
walls, four of which have tension $2\frac{4}{5}$ and two with 
$2\frac{4}{5} \sqrt{2}$, using (\ref{tension}). The junctions allowed
are found by considering the ways that the vacua can be joined
by domain walls. Here there are essentially only two types of
junction, one three-junction and one four-junction, with the angles
involved being found by drawing the vacuum connectivity in the
$W$ space. We know that the tensions of the walls are proportional
to the length of the connection in $W$ space between the vacua (\ref{tension}),
in fact these connections may be used as
a vector diagram for forces. Together the domain walls making up a
junction form a closed polygon when mapped to  the $W$ plane, this
is precisely what is required of a vector diagram of tensions if there
is to be no net force. As an example we consider the junctions of  the
quintic superpotential in fig. \ref{quad_vert}. A triple junction
is calculated by connecting up three vacua in the $W$ plane and
translating this into a closed vector diagram. These vectors then
make up the tensions in the junction, allowing the angles to found
trivially.
One may initially expect the
four-junction to be unstable to producing two triple junctions. This
is not the case as it would require a heavy wall to interpolate
between them which is disfavoured energetically. We undertook a
simulation of this potential, looking for regular tessellations,
testing the stability as before. One pattern which can be formed
uses only the triple junction, 
leading to a familiar `bathroom tiling'
consisting of octagons and squares \cite{grunbaum}. This can be made more intricate
by including the four-junction, a result pictured in
Fig. \ref{oct2_phase}.
We end this catalogue by considering an order seven superpotential,
\mbox{$W(\phi)=\phi-\frac{1}{7}\phi^7$}. This has the possibility
of a rich variety of patterns as there are 5 factorial domain walls
connecting the vacua, which are the sixth roots of unity.  
The walls have three different tensions, 
occuring in  in the ratio 2:$\sqrt{3}$:1.
The triple vertex which has all its walls at a different tension
forms a junction with angles of 120$^\circ$, 150$^\circ$ and 90$^\circ$,
which is easily found using the aforementioned method. One may use this
this to generate a tiling consisting of dodecagons, hexagons and squares,
as illustrated in Fig. \ref{dod_phase}. 
Here we may also generate an attractive tiling using three of the
possible vertices, the triple, quadruple and sextuple junctions.
The result is shown in Fig. \ref{hex_sex_phase}.

In this paper we have indicated the huge variety of
networks that are possible in a relatively simple
field theory.
These networks consist of domains, where the field falls into a
given vacuum, and walls which separate the vacua. 
For stationary solutions we require that each domain
does not 'know' that there is another domain nearby in the same 
vacuum. This necessitates that the domain sizes are larger than the
width of the walls separating them. In fact, violations of the BPS
equation reduce as the wall thickness decreases with respect to the
domain size, approaching the BPS limit as the thickness goes to zero.
It is this same thin defect limit which makes the the string
network's of Sen \cite{sen98} BPS.

One possible area for future study is static, space filling domain
wall networks in three spatial dimensions. At present we see no reason
why such states could not exist, although they may occur more
naturally for a two component complex scalar transforming under 
a natural SU(2) action. 

The networks of domain walls have been shown not to be BPS states,
nevertheless junctions do locally approximately satisfy the BPS 
(or anti-BPS) equation,
with the violation getting smaller as the domain size increases.



{\bf Acknowledgements}: I would like to thank Gary Gibbons for suggesting this project
and for his invaluable comments. Conversations with Nick Manton,
Jesus Moreno and Paul Townsend are also gratefully acknowledged.
This work was supported by PPARC.



\end{document}